\documentclass{vgtc} %
\vgtcinsertpkg

\ifpdf%
  \pdfoutput=1\relax                   %
  \usepackage{graphicx}                %
  \DeclareGraphicsExtensions{.pdf,.png,.jpg,.jpeg} %
\else%
  \usepackage{graphicx}                %
  \DeclareGraphicsExtensions{.eps}     %
\fi%

\graphicspath{{figures/}{}{images/}{./}} %
\usepackage[T1]{fontenc}
\usepackage[utf8]{inputenc}
\usepackage{amsmath}
\usepackage{microtype}                 %
\PassOptionsToPackage{warn}{textcomp}  %
\usepackage{textcomp}                  %
\usepackage{mathptmx}                  %
\usepackage{times}                     %
\usepackage{cite}                      %
\usepackage{tabu}                      %
\usepackage{booktabs}                  %
\usepackage{soul}
\usepackage{cleveref}
\usepackage[font=scriptsize,labelfont=scriptsize]{caption}
\usepackage[labelfont=sf]{subcaption}
\usepackage[dvipsnames]{xcolor}
\usepackage{float}

\onlineid{1165}

\vgtccategory{Research}
\vgtcpapertype{algorithm/technique}

\DeclareRobustCommand{\rev}[1]{{#1}}

\usepackage{array}
\newcommand{\PreserveBackslash}[1]{\let\temp=\\#1\let\\=\temp}
\newcolumntype{C}[1]{>{\PreserveBackslash\centering}p{#1}}
\newcolumntype{R}[1]{>{\PreserveBackslash\raggedleft}p{#1}}
\newcolumntype{L}[1]{>{\PreserveBackslash\raggedright}p{#1}}

\title{Efficient Space Skipping and Adaptive Sampling of Unstructured Volumes Using Hardware Accelerated Ray Tracing\vspace*{-0.2em}}

\author{Nate Morrical$^1$ \;\; Will Usher$^1$ \;\; Ingo Wald$^2$ \;\; Valerio Pascucci$^1$
\\ \parbox{\linewidth}{\scriptsize \centering $^1$SCI Institute, University of Utah \;\; $^2$NVIDIA}}

\abstract{%
    Sample based ray marching is an effective
    method for direct volume rendering of unstructured meshes.
    However, sampling such meshes remains expensive, and
    strategies to reduce the number of samples taken have received relatively
	little attention.
    In this paper, we introduce a method for rendering unstructured 
    meshes using a combination of a coarse spatial acceleration structure and 
    hardware-accelerated ray tracing.
    Our approach enables efficient empty space skipping and adaptive sampling 
    of unstructured meshes, and outperforms a reference ray marcher by up to 7$\times$.
}


\teaser{
  \centering
  \vspace{-2em}
  \resizebox{0.97\textwidth}{!}{
    \includegraphics[height=4.cm]{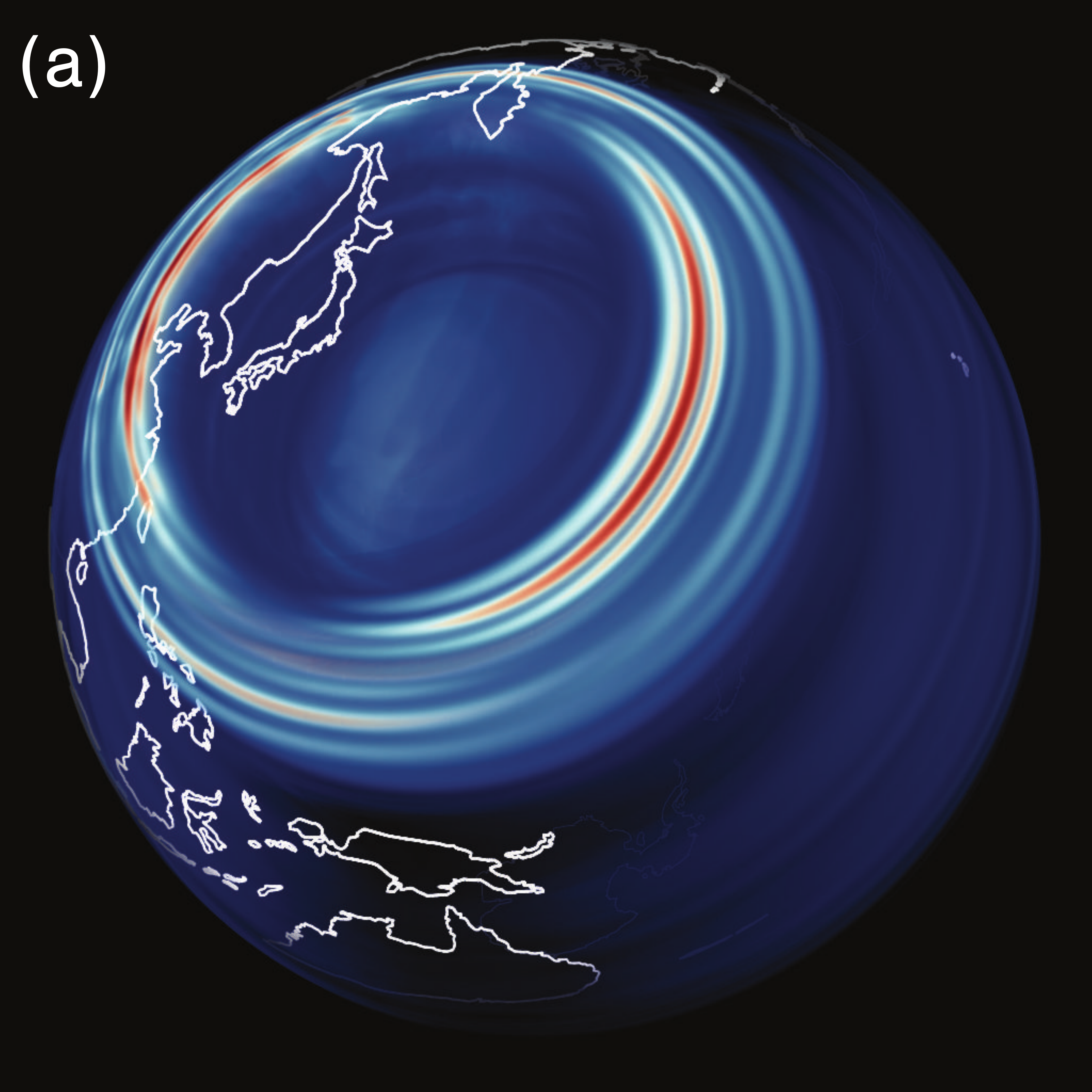}
    \includegraphics[height=4.cm]{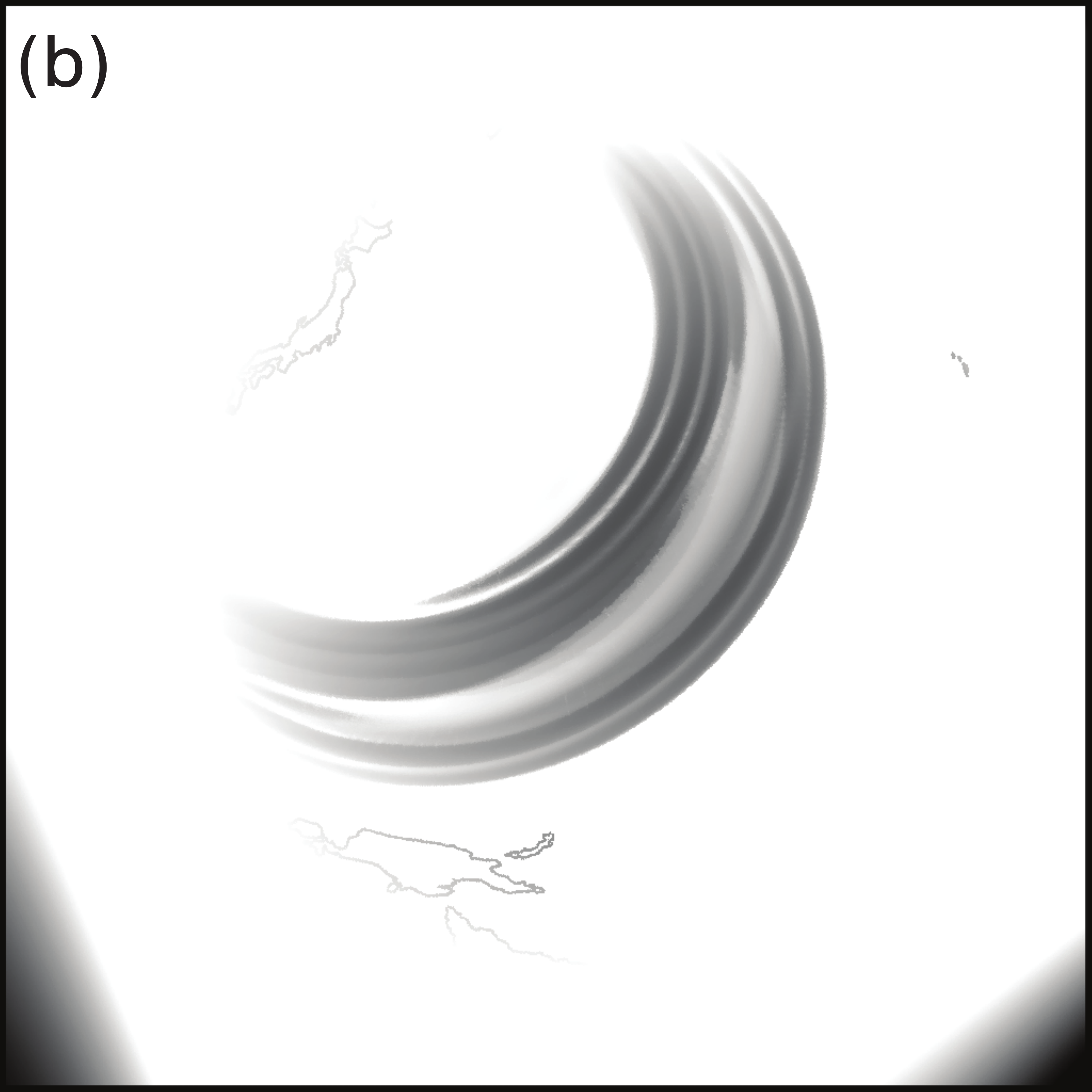}
    \includegraphics[height=4.cm]{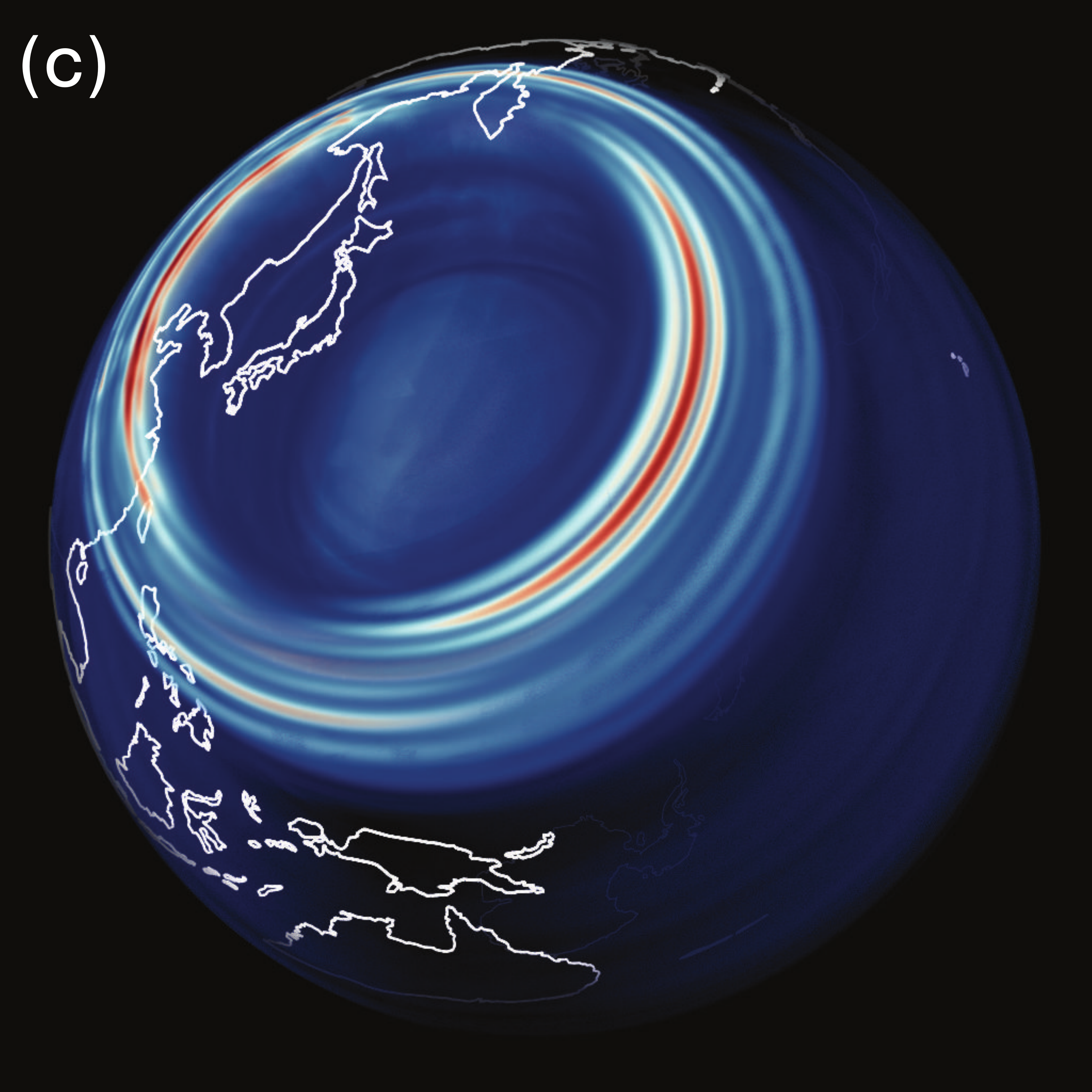}
    \includegraphics[height=4.cm]{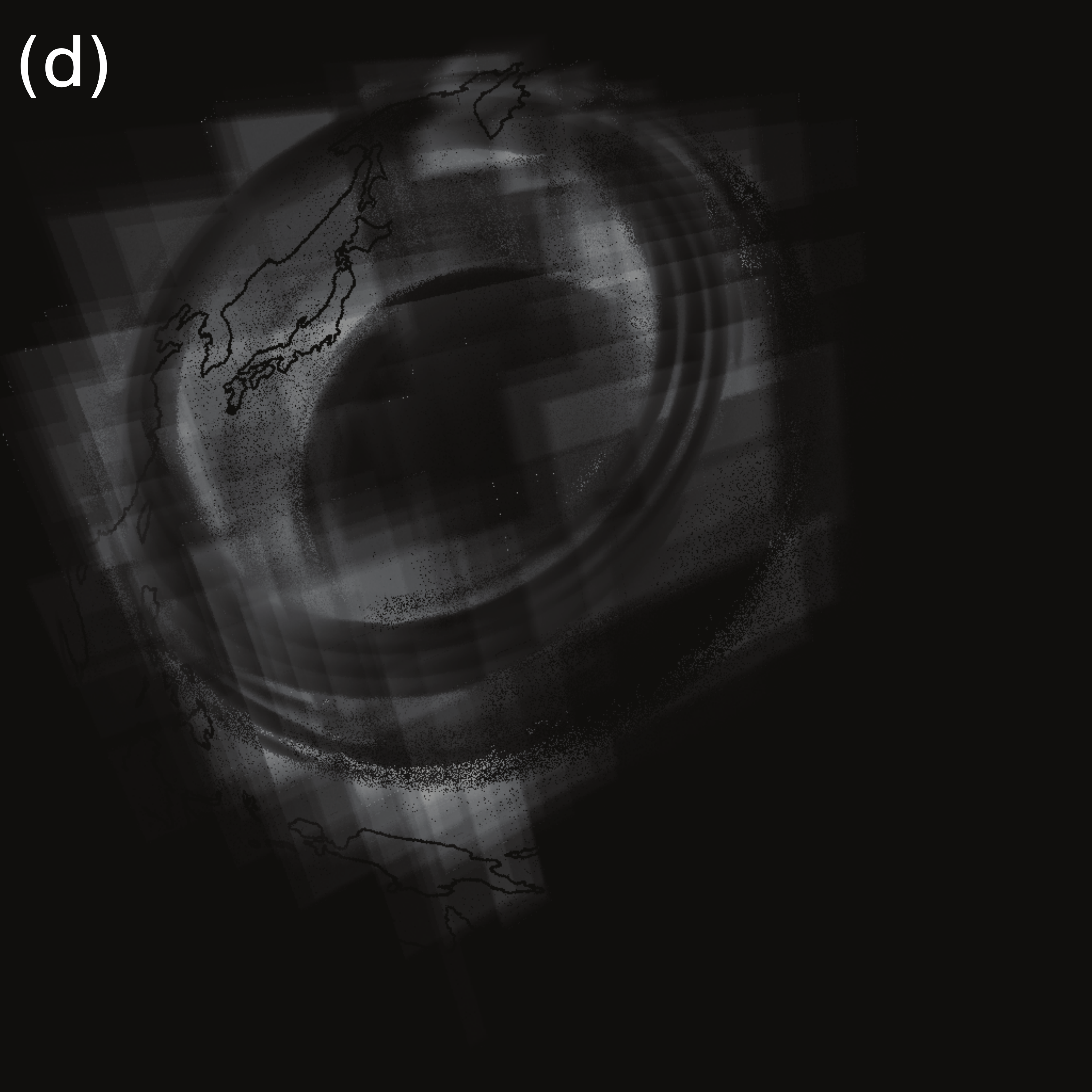}
  }
  \vspace{-.8em}
  \caption{\label{fig:teaser}%
  Performance improvement of our method on the 278~million tetrahedra Japan Earthquake data set.
  (a) A reference volume ray marcher without our method, at 0.9~FPS ($1024^2$ pixels) on an NVIDIA RTX~8000 GPU.
  (b) A heat map of relative cost per-pixel in (a).
  (c) and (d), the same, but now with our space skipping and adaptive sampling method,
  running at 7~FPS (7$\times$ faster).%
  }
  \vspace{-0.5em}
  }

\begin{document}
\firstsection{Introduction}
\maketitle

Direct volume rendering (DVR) is widely used in the scientific visualization community,
enabling scientists to interactively explore their data and form
hypotheses. A standard approach for DVR is raycasting,
where rays are cast through the volume for each pixel and the color and opacity
of the volume is sampled and integrated along each ray to compute an image of the volume.
Interactive ray casting techniques have been demonstrated for both
structured~\cite{parker_interactive_1999,kruger2003acceleration} and
unstructured~\cite{weiler_hardware-based_2003,rathke_simd_2015,nelson_ray-tracing_2006,muigg_interactive_2011,wald_rtx_2019} 
volumes, and map well to the parallel hardware available on modern CPUs~\cite{parker_interactive_1999,wald_ospray_2017,rathke_simd_2015}
and GPUs~\cite{weiler_hardware-based_2003,kruger2003acceleration,nelson_ray-tracing_2006,muigg_interactive_2011,wald_rtx_2019}.

However, when the volume becomes expensive to sample the cost per-ray increases,
limiting interactivity.
For unstructured data, the cost of these samples can be reduced, 
as demonstrated by Wald et al.~\cite{wald_rtx_2019}, by leveraging the ray tracing cores 
available on NVIDIA's Turing GPUs. When used in a naive volume raycaster, their 
approach improved frame rates by $1.5\times$ to $3.8\times$.
To further improve performance, the number of samples taken per-ray must 
be reduced. Numerous methods have been proposed for regular grid volumes,
which roughly fall into two categories: empty-space 
skipping, which avoids sampling fully transparent regions;
and adaptive sampling, which takes fewer samples in
regions containing less interesting data values.
Prior work has employed a range 
of acceleration structures to enable these optimizations, e.g., 
macrocells~\cite{parker_interactive_1999,kruger2003acceleration},
octrees~\cite{boada_multiresolution_2001,gobbetti_single-pass_2008-1,labschutz_jittree_2016,
zimmermann_level--detail_2000,reichl_visualization_2013-1,lamar_multiresolution_1999},
KD-trees~\cite{subramanian_applying_1990,vidal_simple_2008}
and BVHs~\cite{knoll_full-resolution_2011}.

When considering an additional acceleration structure for DVR, the 
performance overhead introduced by building and traversing the structure is of key concern. If the 
overhead incurred by the structure is too high, it can overshadow the performance 
gained from the reduced number of samples taken.
To reduce this overhead, Hadwiger et al.~\cite{hadwiger2017sparseleap} 
proposed SparseLeap, which leverages triangle rasterization hardware to compute 
per-pixel lists of active ray segments by rendering occupancy geometry. These segments 
act as the space skipping acceleration structure in a subsequent render pass. 
\rev{Ganter et al.~\cite{ganter_2019_clustered} recently extended SparseLeap to 
leverage OptiX~\cite{optix} and NVIDIA's ray tracing cores to improve acceleration 
structure build time and use hardware accelerated BVH traversal to
reduce overhead. However, both methods must rebuild the structure on transfer 
function changes, do not consider adaptive sampling, and rely on either an octree 
or a summed area table to build the occupancy geometry, which can result in poor 
adaptivity to an unstructured mesh.}

\begin{figure*}
	\centering
	\vspace{-1.75em}
    \includegraphics[width=\linewidth]{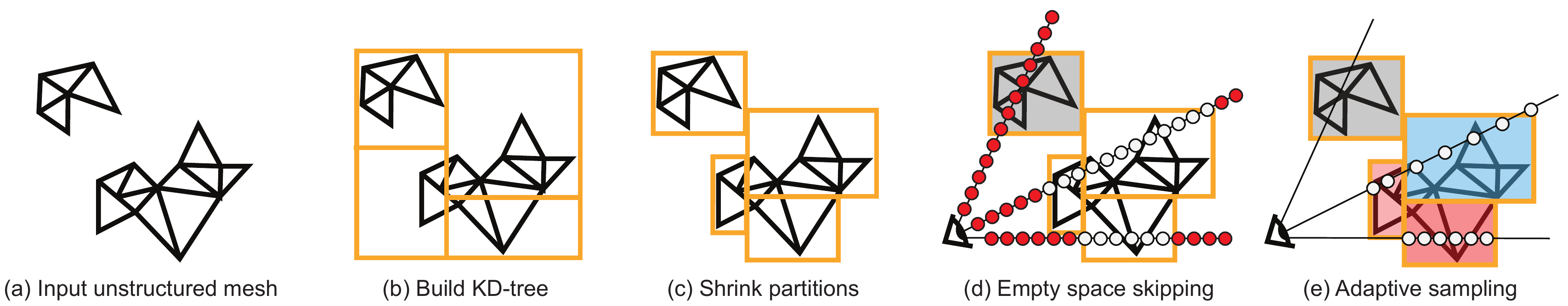}
	\vspace{-1.75em}
	\caption{\label{fig:method}%
	An illustration of our method: (a) Given an unstructured mesh;
	(b) we build a coarse spatial subdivision over the mesh elements to partition them
	into a set of convex, disjoint regions. (c) These partitions are then
	shrunk to tightly fit the contained elements. For each region
	we compute the min/max of the scalar field and the transfer
	function maximum opacity and color variance.
	(d) We use hardware accelerated ray tracing to iterate rays
	through the ``active'' partitions, skipping transparent ones and
	unoccupied space. (e) Within each partition we use the local variance
	to adapt the sampling rate to the underlying data variation, thereby
	taking fewer samples in more uniform regions of the data.}
	\vspace{-1.75em}
\end{figure*}

\rev{In the context of unstructured meshes, relatively little prior work has investigated 
object space adaptive sampling or empty space skipping. The bulk of methods for 
rendering such volumes focus on rasterization methods~\cite{shirley_polygonal_1990,
moreland_fast_2004,callahan_hardware-assisted_2005,maximo_hardware-assisted_2010,callahan_adaptive_2008},
requiring either dynamic level-of-detail strategies~\cite{callahan_2005_k,cook_image_space_2004}
or a volume simplification pre-processing step~\cite{leven_unstructed_textures_2002,garland_2005_quadric,silvia_survey_2005} 
to achieve adaptive sampling.}
\rev{Standard approaches for ray tracing such data~\cite{garrity_raytracing_1990,
bunyk_simple_1997,nelson_ray-tracing_2006,gibeom_accurate_2019} step from cell-to-cell 
along the ray, providing an accurate image at significant cost. Nelson et al.~\cite{nelson_ray-tracing_2006} 
skip individual empty mesh elements, but do not support skipping larger regions at once.}
Prior work has also proposed rasterizing proxy meshes to perform ray tracing on the
GPU~\cite{weiler_hardware-based_2003,muigg_interactive_2011},
but encounter similar adaptive sampling challenges with regard to which proxy 
geometry to dispatch.
Sample based ray casting methods~\cite{rathke_simd_2015} have shown performance 
improvements over both rasterization and cell iteration methods, though have not investigated 
empty space skipping or adaptive sampling.

In this work, we propose new strategies for empty space skipping and adaptive sampling
for sample-based raycasting of unstructured meshes.
\rev{In contrast to prior work, our empty space skipping structure allows skipping
larger regions of space and adapts well to the underlying mesh.
Moreover, we formulate our space skipping structure in a manner
suitable for hardware acceleration, without requiring a rasterizer.
Finally, we propose an intuitive adaptive sampling approach for occupancy geometry 
methods without requiring additional preprocessing.}
Our contributions are:
\begin{itemize}
	\vspace{-0.25em}
	\setlength{\itemsep}{1pt}
	\setlength{\parskip}{1pt}
	\setlength{\parsep}{1pt}
	\item An extension of ``occupancy geometry'' to unstructured data;
	\item A lightweight empty space skipping method for unstructured data, which leverages GPU ray tracing hardware; and
	\item An adaptive sampling approach which provides a bound on error using three intuitive parameters.
\end{itemize}

\section{Method}
\rev{Unstructured meshes pose unique challenges which must be overcome
to implement an effective empty space skipping and adaptive sampling strategy.
Such meshes are often refined by the simulation in areas of interest, and
the method must adapt to this non-uniformity, where the size and
distribution of the elements is not known a priori.
Furthermore, as the mesh is not guaranteed to be convex, the
method must account for three categories of coarser regions: 
unoccupied regions containing no elements,
regions containing entirely transparent elements, and regions
where the data value does not vary significantly across the elements.}

In our approach, we first partition 
the volume into a set of convex, disjoint regions (\Cref{sec:partition_gen}),
which are shrunk to tightly fit the contained mesh elements (\Cref{sec:partition_refine}).
We then compute metadata about the elements contained in these partitions,
which we use to guide both the empty-space skipping (\Cref{sec:space_skipping})
and adaptive sampling (\Cref{sec:adaptive_sampling}).
Finally, we use the Turing GPU's new ray tracing hardware to accelerate traversal of these partitions
for empty space skipping, and our adaptive sampling method to reduce the number 
of samples taken in each partition.
An overview of our method is shown in~\Cref{fig:method}.

\subsection{Partition Generation}
\label{sec:partition_gen}

To partition the mesh elements, %
we use the leaves of a spatial KD-tree (\Cref{fig:method}b), which are 
convex, disjoint, and adaptive.
The disjoint, non-overlapping property ensures that a ray will exit one partition 
before entering the next, and the convexity property ensures that the ray 
will enter each partition only once.
As the elements in the mesh are unlikely to be uniform in size or distribution,
the adaptivity in partition size provided by the KD-tree is desirable
\rev{over a more fixed structure
(e.g., grids~\cite{ganter_2019_clustered}, octrees~\cite{hadwiger2017sparseleap})} to ensure
a roughly even distribution of rendering cost for the generated partitions.
We note that it is possible for a tetrahedron (or other mesh element)
to appear in more than one leaf node, and thus partition.
In this case, rays entering a given partition
will only sample the portion of the element contained inside the partition's bounds.

For each partition we then compute and store the scalar field value range for the contained elements.
For those elements which are only partly contained in the partition we still include the value
range of the entire element for simplicity.
When the transfer function is changed we apply it across the stored
range of field values contained in the partition, to determine the maximum opacity and color
variance of the partition. The complexity of the opacity and variance computation is linear
in the number of values in the transfer function, and is independent
of the number of elements contained in the partition. As there are likely far
fewer values in the transfer function than elements in the mesh, this provides
better responsiveness to user changes to the transfer function.
The variance computation is
parallelized across the partitions, allowing for faster updates.
The per-partition variance values are then normalized relative to the minimum
and maximum variances over all partitions, to allow computing consistent per-partition
sampling rates.

\subsection{Partition Refinement}
\label{sec:partition_refine}
Although the KD-tree leaves provide the desired convex and disjoint partitioning
of the volume, the bounds of a leaf do not necessarily tightly bound the
contained elements (see~\Cref{fig:method}b). As such, the initially computed partitions 
may contain large regions of unoccupied space, especially in the case of non-convex,
hollow, or non-grid aligned meshes.
To provide tighter bounds, we shrink the bounding box of each partition to fit the bounding 
box of the contained elements (\Cref{fig:method}c).
The shrunk bounds is the intersection
of the bounds of the contained elements and the original leaf node bounds,
to ensure it does not expand out of the original leaf node.
\subsection{Empty Space Skipping}
\label{sec:space_skipping}
Given the bounds of the partitions, we can leverage hardware accelerated ray tracing 
to accelerate intersection tests against the partitions to find
ray entry and exit points. The RT cores available in Turing GPUs support both hardware accelerated BVH traversal
and ray-triangle intersection tests. To fully utilize the available hardware, we first
tessellate the partition bounding boxes, then construct an OptiX~\cite{optix} BVH over the generated
triangles. This BVH need only be rebuilt when the underlying partition geometry
changes, and is not tied to the transfer function.

To find the entry and exit points of the ray in some partition we use OptiX to 
trace rays against the partition geometry; first with back-face culling enabled 
to find the entry point, then from the entry point with front-face culling enabled to find exit point. 
The $t$ range along the ray between $t_{\text{enter}}$ and $t_{\text{exit}}$ is thus the range to 
integrate the volume over to sample the partition. If the ray intersects a completely transparent partition
it is skipped and we advance the ray to find the next partition.
If no partition is found, the ray is terminated and
the computed color and opacity is composited with the background and written
to the framebuffer.

To advance to the next partition we set the ray's $t_{\min}$ to
$t_{\text{exit}} - \epsilon$. We apply a small offset back by $\epsilon$ to
allow for intersection with potentially coplanar partition boundary faces.
From this new start point we then find the ray's entry and exit points with the
next partition as before.

\subsection{Adaptive Sampling}
\label{sec:adaptive_sampling}
When ray marching through a partition, we want to reduce the number of samples
taken to better match the local partition variance. In partitions with
relatively similar colors, the variance is low, and a correspondingly low number
of samples can be taken. Regions with wider color variation require more samples 
to preserve accuracy.

To adaptively sample each partition we use the transfer function variance for the partition
computed in \Cref{sec:partition_gen} to select the step size for the ray marching process.
The step size is computed using an intuitive equation which allows
users to place an upper bound on the tolerable maximum step size, and thus error;
and control how quickly the algorithm transitions from high to low quality
sampling based on the partition's variance.
Given a minimum step size $s_1$ to use for the highest quality sampling and the maximum
step size $s_2$ to use for the lowest quality sampling, we compute the
step size for a partition with normalized variance $\sigma^2$ using
${s = \max\left(s_1\ + (s_2 - s_1) | \min(\sigma,\ 1) - 1 |^{p},\ s_1\right)}$.

The final user controllable parameter is $p$, referred to as the \textit{adaptive power},
which allows the user to tune how quickly the algorithm will transition to lower
quality sampling in medium variance partitions. We restrict that $p\ge1$, as this
lower bound corresponds to a simple linear interpolation between $s_1$ and $s_2$.

With this equation the user can easily tune the sampling quality to produce an
acceptable image at some desired frame rate. If the user wants
a lower quality image at a higher frame rate they can increase $s_1$ and $s_2$,
or for a more expensive but higher-quality image, decrease both values.
If desired, the adaptive sampling can be disabled entirely by setting $s_1 = s_2$.
If the user finds too few
samples are taken in partitions with medium variance they can increase $p$
to bias $s$ towards $s_1$ in these partitions. Similarly,
to improve frame rate at the cost of error in medium variance partitions $p$
can be decreased, to bias $s$ towards $s_2$.

Given the sampling step size for a partition
we integrate the ray front to back through the partition, using the \texttt{rtx-shared-faces}
point query kernel described by Wald et al.~\cite{wald_rtx_2019} to
sample at points along the ray.
To ensure correct opacity when compositing partitions integrated at different step
sizes we use an opacity correction term~\cite{engel_real-time_2006}. Given
the current sample's opacity $\alpha$ we compute the corrected opacity as 
$\tilde{\alpha} = 1 - (1 - \alpha)^{s / s_1}$.
Finally, we perform early ray termination if the ray becomes opaque.

\section{Evaluation}
\begin{figure}[t!]
  \centering
  \begin{tabular}{C{.95\columnwidth}}
    \toprule
	  \scriptsize{Jets, 12M~tets (vertex centered data)} \\
	  \midrule
  \end{tabular}\\
	\begin{subfigure}{0.28\columnwidth}
		\centering
		\includegraphics[width=\linewidth]{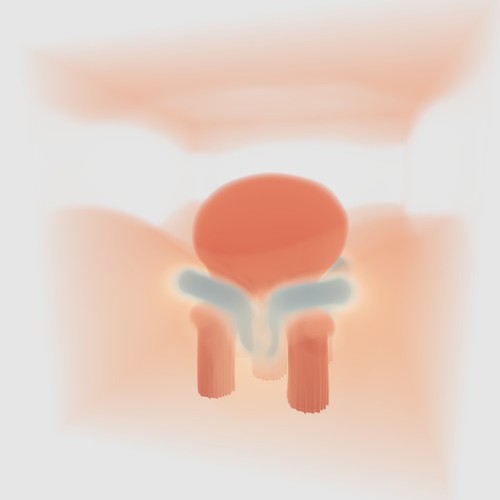}
	\vspace{-1.7em}
		\caption{Reference, 4.8~FPS}
	\end{subfigure}
	\begin{subfigure}{0.28\columnwidth}
		\centering
		\includegraphics[width=\linewidth]{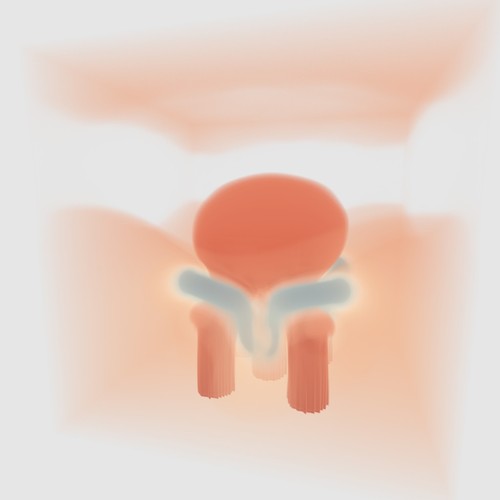}
	\vspace{-1.7em}
		\caption{Adaptive, 16.7~FPS}
	\end{subfigure}
	\begin{subfigure}{0.28\columnwidth}
		\centering
		\includegraphics[width=\linewidth]{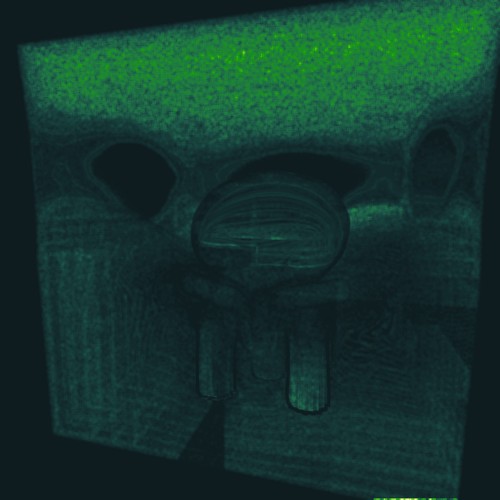}
	\vspace{-1.7em}
		\caption{SSIM, .997}
	\end{subfigure}
  \begin{tabular}{C{.95\columnwidth}}
    \toprule
	  \scriptsize{Agulhas Current, 35M~tets (cell centered data)} \\
    \midrule
  \end{tabular}\\
	\begin{subfigure}{0.28\columnwidth}
		\centering
		\includegraphics[width=\linewidth]{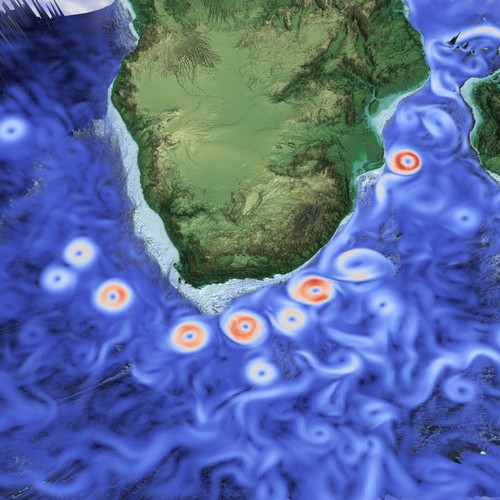}
	\vspace{-1.7em}
		\caption{Reference, 14~FPS}
	\end{subfigure}
	\begin{subfigure}{0.28\columnwidth}
		\centering
		\includegraphics[width=\linewidth]{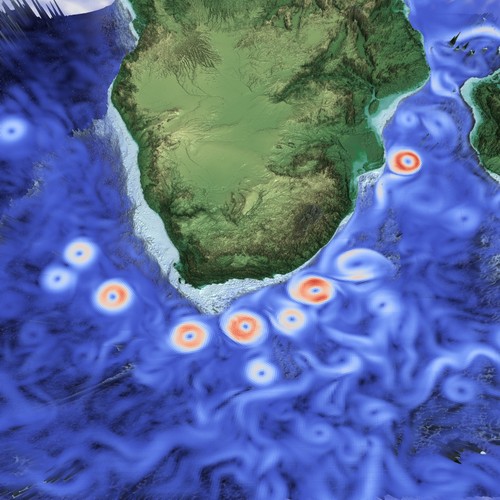}
	\vspace{-1.7em}
		\caption{Adaptive, 48~FPS}
	\end{subfigure}
	\begin{subfigure}{0.28\columnwidth}
		\centering
		\includegraphics[width=\linewidth]{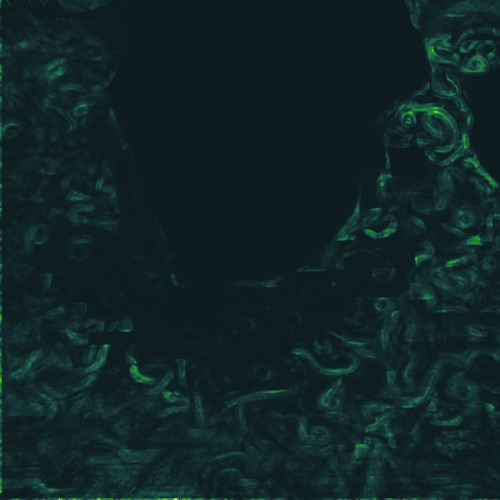}
	\vspace{-1.7em}
		\caption{SSIM, .98}
	\end{subfigure}
  \begin{tabular}{C{.95\columnwidth}}
    \toprule
	  \scriptsize{Japan Earthquake, 278M~tets (vertex centered data)} \\
    \midrule
  \end{tabular}\\
	\begin{subfigure}{0.28\columnwidth}
		\centering
		\includegraphics[width=\linewidth]{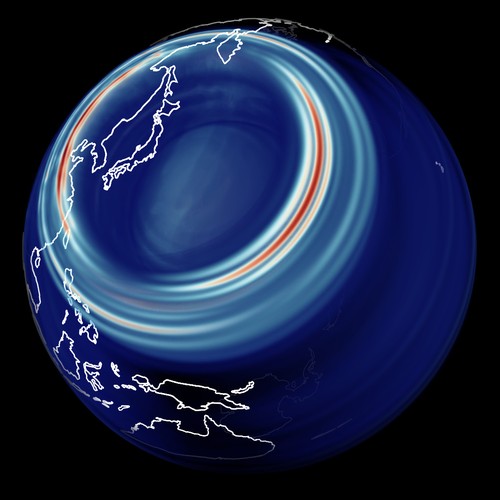}
	\vspace{-1.7em}
		\caption{Reference, 0.9~FPS}
	\end{subfigure}
	\begin{subfigure}{0.28\columnwidth}
		\centering
		\includegraphics[width=\linewidth]{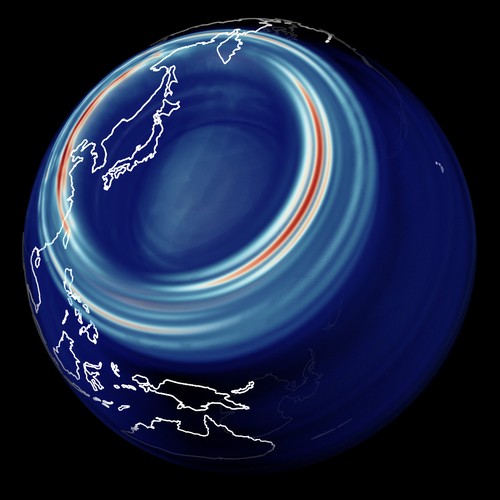}
	\vspace{-1.7em}
		\caption{Adaptive, 7~FPS}
	\end{subfigure}
	\begin{subfigure}{0.28\columnwidth}
		\centering
		\includegraphics[width=\linewidth]{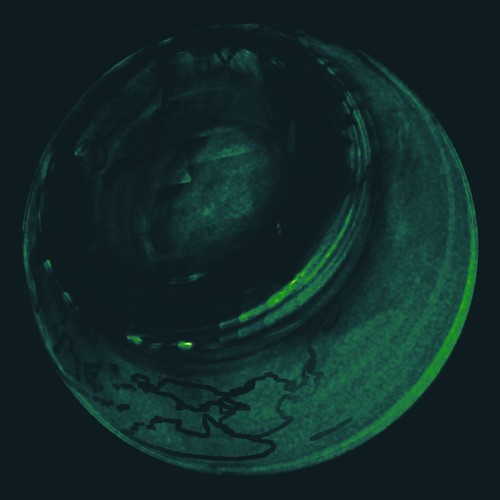}
	\vspace{-1.7em}
		\caption{SSIM, .98}
	\end{subfigure}
  \begin{tabular}{C{.95\columnwidth}}
    \toprule
	  \scriptsize{Deep Water Asteroid Impact, 366M~tets (cell centered data)} \\
    \midrule
  \end{tabular}\\
	\begin{subfigure}{0.28\columnwidth}
		\centering
		\includegraphics[width=\linewidth]{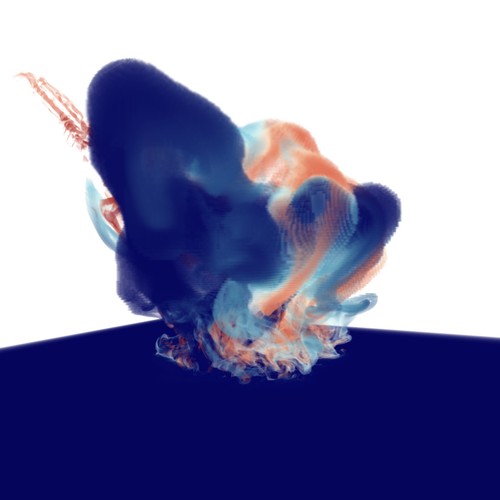}
	\vspace{-1.7em}
		\caption{Reference, 4~FPS}
	\end{subfigure}
	\begin{subfigure}{0.28\columnwidth}
		\centering
		\includegraphics[width=\linewidth]{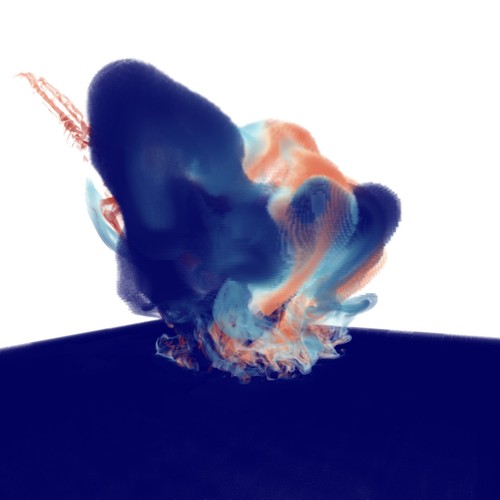}
	\vspace{-1.7em}
		\caption{Adaptive, 14~FPS}
	\end{subfigure}
	\begin{subfigure}{0.28\columnwidth}
		\centering
		\includegraphics[width=\linewidth]{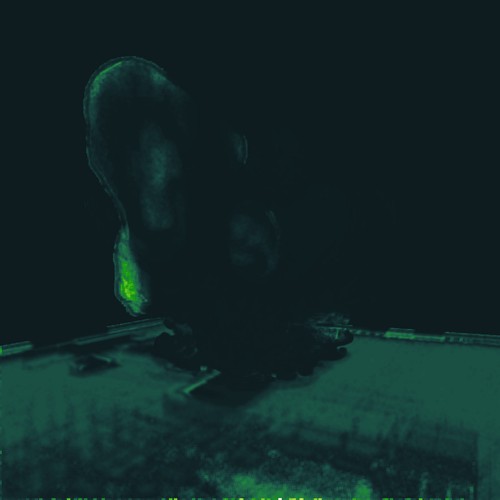}
	\vspace{-1.7em}
		\caption{SSIM, .97}
	\end{subfigure}
	\vspace{-0.75em}
	\caption{\label{fig:benchmark_imgs}%
	Quality and performance
	comparisons of our method against a reference volume ray
	marcher~\cite{wald_rtx_2019}, using
	representative views and transfer functions. For comparable
	image quality our method performs roughly $3-7\times$
	faster, achieving its greatest speedup in the most irregular
	data set (Japan Earthquake). More aggressive quality
	settings can yield higher speedups, at the cost
	of image quality. For larger images, please see the supplemental material.}
	\vspace{-0.25em}
\end{figure}

We evaluate our approach using four tetrahedral mesh volumes (\Cref{fig:benchmark_imgs})
covering a range of data set sizes, on an NVIDIA RTX~8000 GPU, primarily due to its 
large memory capacity. Our renderer is implemented using OptiX 6 and CUDA 10.
For the Jets, Agulhas Current and Japan Earthquake datasets we set $p = 2$,
on the Deep Water Asteroid Impact we set $p = 6$.
We first evaluate the performance gains provided by our empty space skipping
method (\Cref{sec:eval_skipping}), after which we combine it with our adaptive sampling method
and evaluate the two in combination (\Cref{sec:eval_adaptive}).
Finally, we measure the overhead incurred by the two methods
in \Cref{sec:eval_overhead}.

\begin{figure*}
	\centering
	\vspace{-1.75em}
	\includegraphics[width=\linewidth]{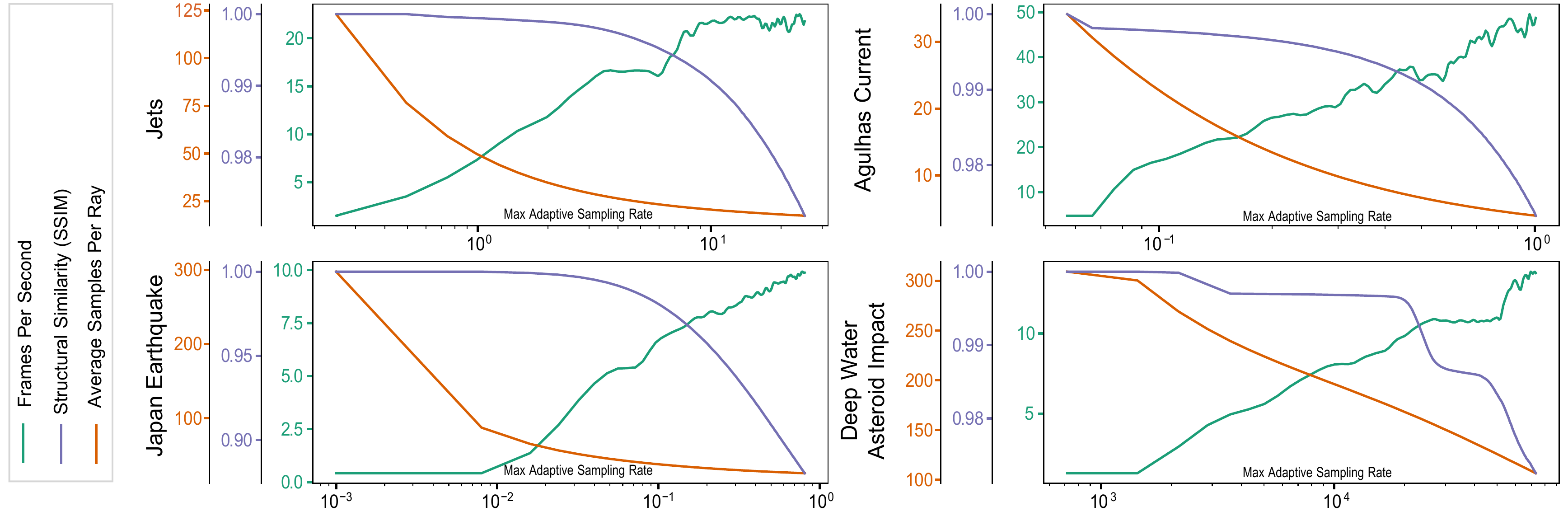}
	\vspace{-2.5em}
	\caption{\label{fig:quality_vs_speed}%
	The effect of increasing the maximum step size (tolerable error) on
	rendering performance, samples taken and image quality for each data set.
	As expected, image quality decreases as the maximum step size
	is increased, though remains high-quality (SSIM $\ge$ 0.97) even
	when taking just a fraction of the original samples required. Using
	our approach users can tune the error incurred as desired to achieve
	improved frame rates for unstructured volume rendering.}
	\vspace{-1.5em}
\end{figure*}

\subsection{Space Skipping Improvements}
\label{sec:eval_skipping}
\begin{figure}[t]
	\centering
\begin{subfigure}{0.3\columnwidth}
	\centering
	\includegraphics[width=\linewidth]{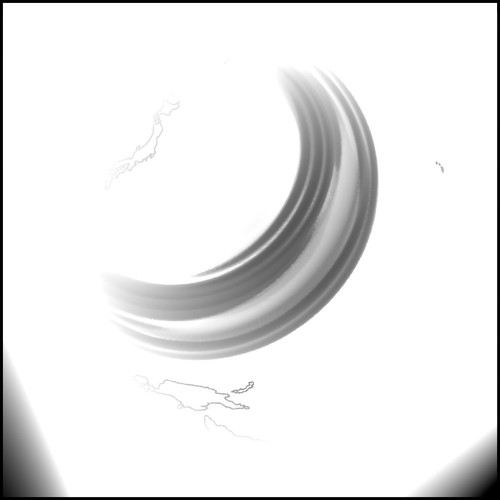}
	\caption{Reference}
\end{subfigure}
\begin{subfigure}{0.3\columnwidth}
	\centering
	\includegraphics[width=\linewidth]{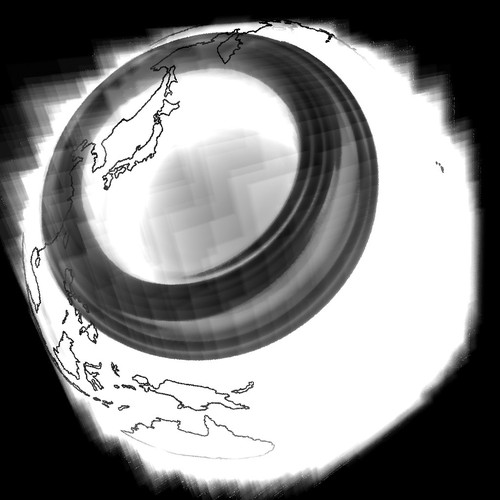}
	\caption{\label{fig:space_skipping_samples}%
	Space skipping only.}
\end{subfigure}
\begin{subfigure}{0.3\columnwidth}
	\centering
	\includegraphics[width=\linewidth]{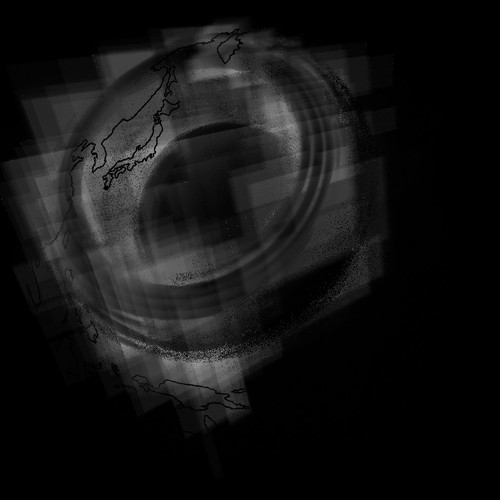}
	\caption{\label{fig:adapt_skip_samples}%
		Plus adapt.\ sampling.}
\end{subfigure}
\vspace{-.8em}
\caption{\label{fig:sample_reduction_imgs}%
	A heatmap of the samples taken per-pixel compared to the reference (a),
	when using (b) just space skipping, and (c) space skipping plus adaptive sampling.
	(a)~The reference takes a large number of samples for most pixels, except a few where
	early ray termination occurs.
	(b)~Space skipping avoids unoccupied and fully transparent partitions, though
	takes many samples in visible partitions.
	(c)~Adaptive sampling reduces samples taken in visible low-variance regions,
	reducing samples while providing similar image quality.}
	\vspace{-0.5em}
\end{figure}

Empty space skipping is able to reduce the samples taken per-ray in two ways: 
by skipping regions outside the volume, and by skipping 100\% transparent partitions. 
These regions can be skipped, since they do not contribute to the final 
image (\Cref{fig:space_skipping_samples}).

With regard to the former, we observe a negligible performance
improvement when only skipping unoccupied space compared to naively
taking samples potentially outside the volume. The volumes we conduct
our evaluation on are relatively dense, providing little unoccupied space
to skip in the first place. Moreover, it is likely that the hardware accelerated
BVH traversal performed when querying a point is
able to quickly determine the point is outside the volume and terminate,
incurring little cost per-sample.

The performance improvement provided in the latter case, skipping 100\% transparent
partitions, is highly dependent on the transfer function chosen by the user (\Cref{fig:impact-of-xf}).
When using a relatively ``binary'' transfer function, where background regions
are made entirely transparent, a large number of partitions can be discarded,
yielding a significant performance improvement. However, if these background
regions are made slightly opaque it is no longer possible to discard them,
and relatively few empty partitions can be skipped, thereby limiting the performance
improvement which can be achieved.

\subsection{Adaptive Sampling Improvements}
\label{sec:eval_adaptive}
\begin{figure}[t]
	\centering
	\vspace{-0.0em}
	\begin{tabular}{cc}
	  \includegraphics[width=.3\columnwidth]{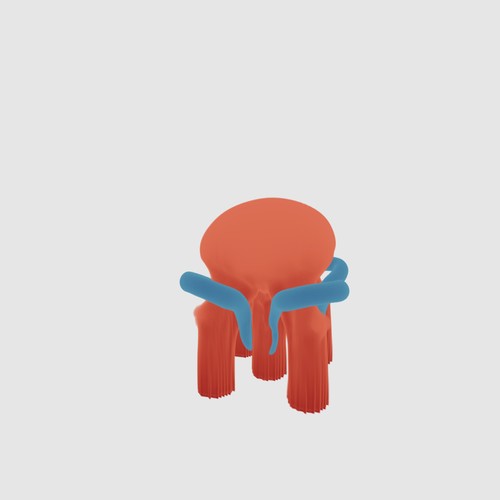}
	  &
	  \includegraphics[width=.3\columnwidth]{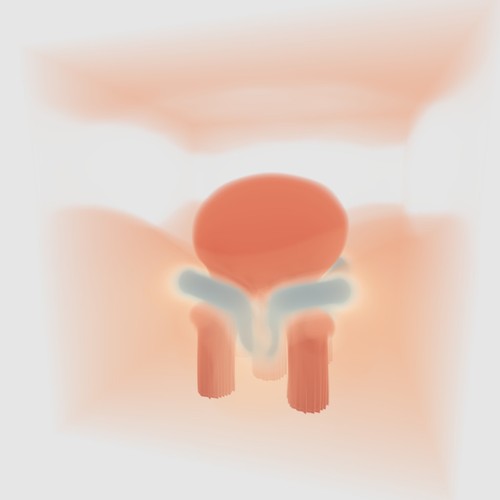}
	  \\
		{\tiny\textsf{Reference: 4~FPS}\vspace{-0.22em}}
	  &
		{\tiny\textsf{Reference: 4~FPS}\vspace{-0.22em}}
	  \\
		{\tiny\textsf{Space skipping only: 20~FPS}\vspace{-0.22em}}
	  &
		{\tiny\textsf{Space skipping only: 5~FPS}\vspace{-0.22em}}
	  \\
	  	{\tiny\textsf{\rev{Adaptive only: 10~FPS}}\vspace{-0.22em}}
	  &
		{\tiny\textsf{\rev{Adaptive only: 12~FPS}}\vspace{-0.22em}}
	  \\
		{\tiny\textsf{Together: 24~FPS}}
	  &
		{\tiny\textsf{Together: 13~FPS}}
	  \end{tabular}
	  \vspace{-0.75em}
	\caption{\label{fig:impact-of-xf}%
	  Empty space skipping works best for ``binary'' transfer functions,
	  where parts of the volume are 100\% transparent
	  (left); but on its own breaks down if regions are not 100\%
	  transparent (right). Adaptive sampling is able to reduce the
	  sampling rate in these semitransparent low-variance regions,
	  improving performance for such cases.}
	  \vspace{-0.5em}
\end{figure}

Our adaptive sampling approach can provide significant performance improvements
by reducing samples taken in low variance partitions (\Cref{fig:adapt_skip_samples}),
with little sacrifice in image quality.
When combined with our empty space skipping approach, adaptive sampling
improves performance in semitransparent low-variance regions (see~\Cref{fig:impact-of-xf}).
For high-quality rendering using both methods in combination
we find significant performance improvements. On the Jets, Agulhas Current and Deep
Water Asteroid Impact we achieve a roughly 3.5$\times$ improvement in rendering
performance; on the Japan Earthquake we achieve a 7.8$\times$
improvement; in all cases SSIM~$\ge$~0.97 (\Cref{fig:benchmark_imgs}).

In~\Cref{fig:quality_vs_speed} we evaluate the impact on frame rate, samples taken and image
quality as the tolerable maximum step size $s_2$ is increased. As $s_2$ is increased,
the adaptive sampling can take larger steps in low and medium variance regions,
reducing samples taken and thereby improving performance, at the cost of image quality.
At the extreme end we find that even when taking $1/3$ or fewer samples than the reference,
a tolerable medium to high-quality image can still be provided.
Moreover, significant reductions in samples taken, and thus increases in performance,
can be achieved with little perceivable impact on image quality.
For volumes with more expensive sampling, e.g., higher-order interpolants or
non-hardware accelerated tet-mesh sampling, the performance improvement achieved
by reducing the number of samples taken is likely to be even greater.

\subsection{Acceleration Structure Overhead}
\label{sec:eval_overhead}
In our evaluation we found the added work of intersecting rays with the partition
boundaries is negligible. With relatively few generated partitions, even for
large data sets, and hardware accelerated ray tracing used to intersect these
partitions, there is little cost incurred to traverse them.
For example, on the Japan Earthquake, which contains the largest number of partitions (4725),
we find tracing rays through these partitions takes just 3ms.

\section{Limitations, Future Work, and Conclusion}
We have presented new strategies for leveraging empty space skipping and adaptive
sampling in the context of volume rendering unstructured meshes.
Our method significantly reduces the number of samples
required per-pixel, improving performance without sacrificing image quality.
Our adaptive sampling
method exposes an intuitive set of parameters to users, allowing
them to easily control the trade off between performance and image quality.
Furthermore, our approach is able to leverage the new ray tracing hardware available on
recent GPUs and incurs little overhead as a result.

Although we have demonstrated significant performance improvements when using our
method, it is not without limitations.
First, as with other adaptive sampling approaches we find diminishing
returns as the sampling rate becomes very low.
\rev{Second, our adaptive sampling is only based on the transfer function, and
would require additional metadata per partition to account for
gradient shading or scattering.}
\rev{Third, our occupancy geometry must be rebuilt on mesh geometry changes,
which would impact performance on both time series data as well as runtime level-of-detail strategies.}
On large datasets such as the Deep Water Asteroid Impact,
we encountered numerical precision issues when traversing the partitions.
These precision issues may be addressed by adapting our epsilon offset to account for the
data set size, or using a custom higher-precision intersection test for the partition geometry.
Finally, although the median split KD-tree provides a reasonable spatial partitioning,
taking into account the underlying scalar field may provide better results. For example,
on the Deep Water Asteroid Impact the water is split into multiple partitions; however,
the field is uniform across the region and a single partition would suffice.

In future work, it would be valuable to explore a method for automatically
selecting the adaptive sampling parameters based on some runtime refinement
to provide a desired image quality, instead of requiring users to
manually tune the sampling parameters.
Although we have evaluated our method in a GPU raycaster to
leverage hardware accelerated ray tracing, it could
translate well onto the CPU using Embree~\cite{wald_embree_2014}.
While we have only evaluated our method on linearly interpolated
tetrahedral meshes, a similar approach may work well for other unstructured
volumes, adaptive mesh refinement (AMR) volumes, and higher order interpolants.

\acknowledgments{%
The Agulhas is courtesy Dr. Niklas Röber (DKRZ); 
the Japan Earthquake is courtesy of Carsten Burstedde, Omar Ghattas, James R. Martin,
Georg Stadler, and Lucas C. Wilcox (ICES, the UT
 Austin) and Paul Navrátil and Greg Abram (TACC);
the Deep Water Asteroid Impact is courtesy of John Patchett and Galen Gisler of LANL.
Hardware for development and testing was graciously provided by NVIDIA Corp.
This work is supported in part by NSF: CGV Award: 1314896, NSF:IIP Award: 1602127, NSF:ACI Award: 1649923, DOE/SciDAC DESC0007446, CCMSC DE-NA0002375 and NSF:OAC
Award: 1842042.
}

\newpage
\bibliographystyle{abbrv-doi}

\bibliography{ms}

\end{document}


\begin{figure*}
    \vspace{-4.8em}
    \centering
    \begin{tabular}{C{.95\linewidth}}
      \toprule
      Jets, 12M~tets (vertex centered data) \\
        \midrule
    \end{tabular}\\
      \begin{subfigure}{0.32\linewidth}
          \centering
          \includegraphics[width=\linewidth]{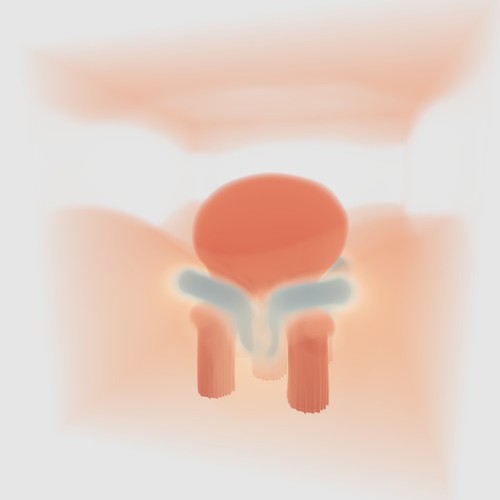}
      \vspace{-1.7em}
          \caption{Reference, 4.8~FPS}
      \end{subfigure}
      \begin{subfigure}{0.32\linewidth}
          \centering
          \includegraphics[width=\linewidth]{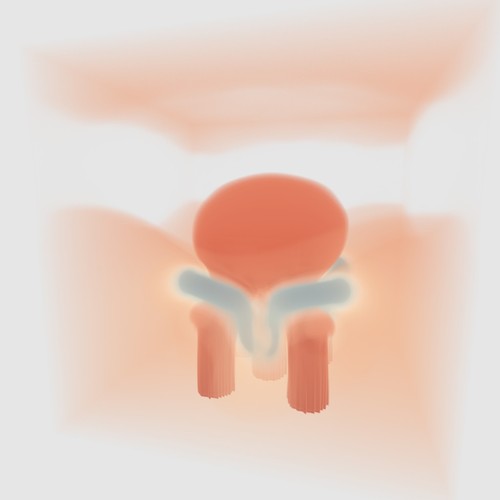}
      \vspace{-1.7em}
          \caption{Adaptive, 16.7~FPS}
      \end{subfigure}
      \begin{subfigure}{0.32\linewidth}
          \centering
          \includegraphics[width=\linewidth]{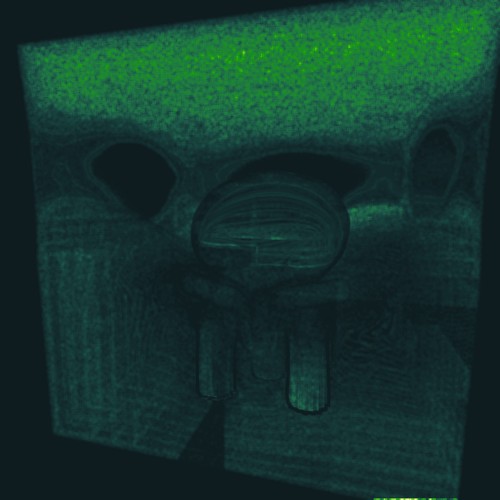}
      \vspace{-1.7em}
          \caption{SSIM, .997}
      \end{subfigure}
    \begin{tabular}{C{.95\linewidth}}
      \toprule
      Agulhas Current, 35M~tets (cell centered data) \\
      \midrule
    \end{tabular}\\
      \begin{subfigure}{0.32\linewidth}
          \centering
          \includegraphics[width=\linewidth]{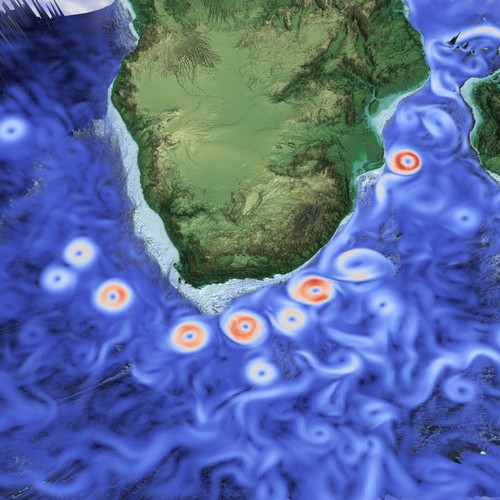}
      \vspace{-1.7em}
          \caption{Reference, 14~FPS}
      \end{subfigure}
      \begin{subfigure}{0.32\linewidth}
          \centering
          \includegraphics[width=\linewidth]{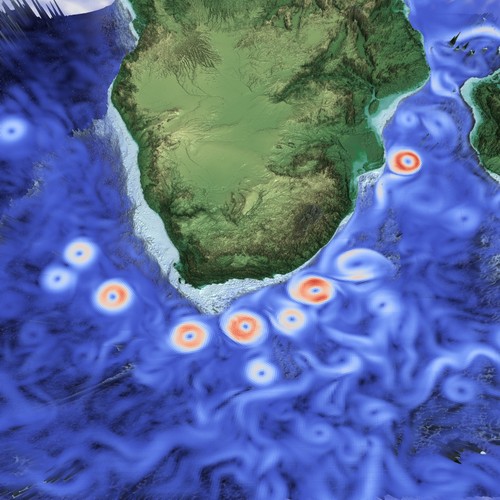}
      \vspace{-1.7em}
          \caption{Adaptive, 48~FPS}
      \end{subfigure}
      \begin{subfigure}{0.32\linewidth}
          \centering
          \includegraphics[width=\linewidth]{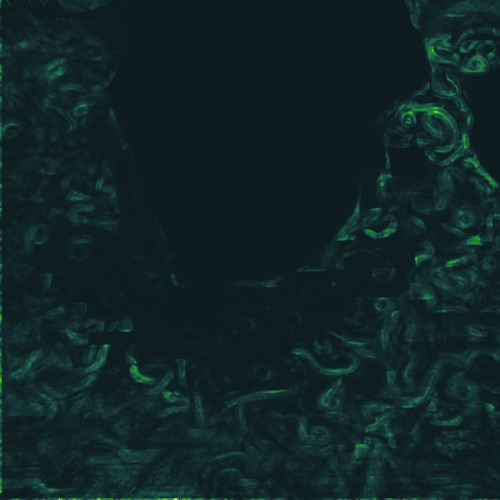}
      \vspace{-1.7em}
          \caption{SSIM, .98}
      \end{subfigure}
    \begin{tabular}{C{.95\linewidth}}
      \toprule
      Japan Earthquake, 278M~tets (vertex centered data) \\
      \midrule
    \end{tabular}\\
      \begin{subfigure}{0.32\linewidth}
          \centering
          \includegraphics[width=\linewidth]{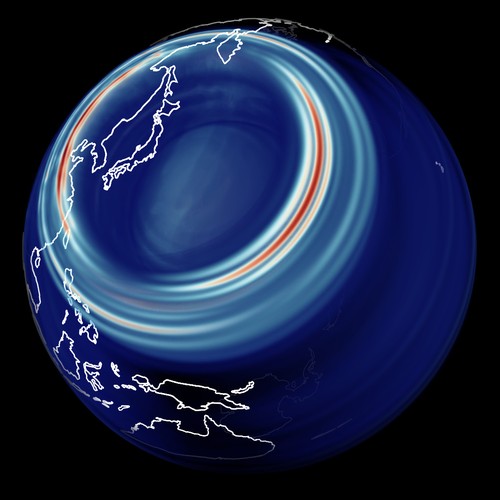}
      \vspace{-1.7em}
          \caption{Reference, 0.9~FPS}
      \end{subfigure}
      \begin{subfigure}{0.32\linewidth}
          \centering
          \includegraphics[width=\linewidth]{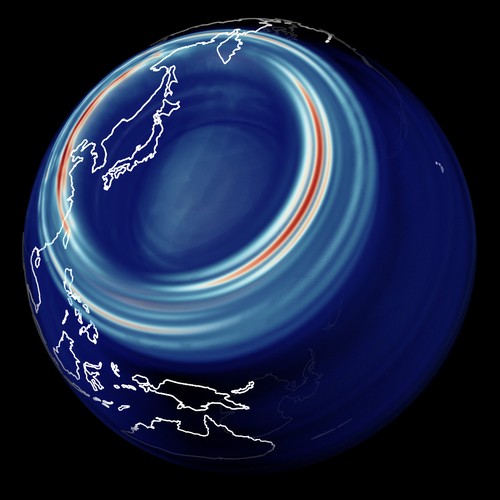}
      \vspace{-1.7em}
          \caption{Adaptive, 7~FPS}
      \end{subfigure}
      \begin{subfigure}{0.32\linewidth}
          \centering
          \includegraphics[width=\linewidth]{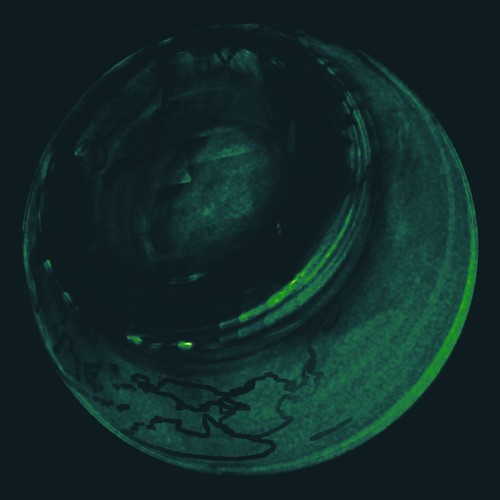}
      \vspace{-1.7em}
          \caption{SSIM, .98}
      \end{subfigure}
    \begin{tabular}{C{.95\linewidth}}
      \toprule
      Deep Water Asteroid Impact, 366M~tets (cell centered data) \\
      \midrule
    \end{tabular}\\
      \begin{subfigure}{0.32\linewidth}
          \centering
          \includegraphics[width=\linewidth]{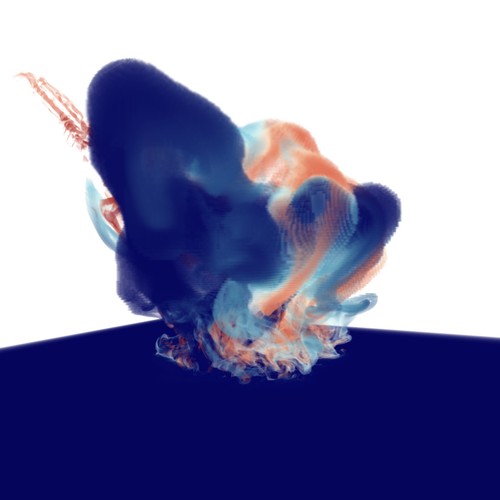}
      \vspace{-1.7em}
          \caption{Reference, 4~FPS}
      \end{subfigure}
      \begin{subfigure}{0.32\linewidth}
          \centering
          \includegraphics[width=\linewidth]{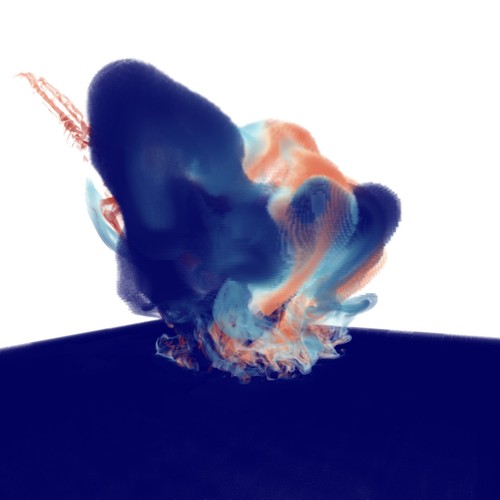}
      \vspace{-1.7em}
          \caption{Adaptive, 14~FPS}
      \end{subfigure}
      \begin{subfigure}{0.32\linewidth}
          \centering
          \includegraphics[width=\linewidth]{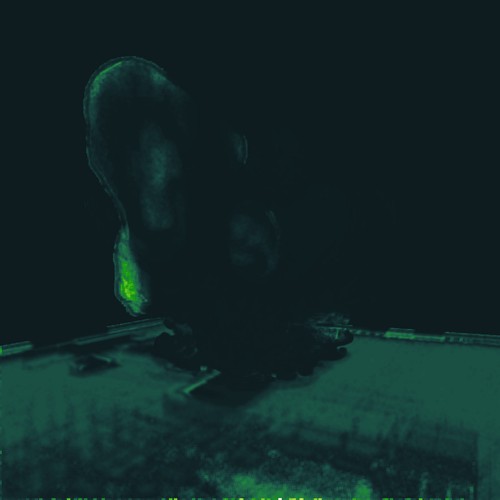}
      \vspace{-1.7em}
          \caption{SSIM, .97}
      \end{subfigure}
      \vspace{-0.5em}
  \end{figure*}